# The Manipulation Problem: Conversational AI as a Threat to Epistemic Agency

Louis B. Rosenberg

Unanimous AI

Louis@Unanimous.ai

The technology of Conversational AI has made significant advancements over the last eighteen months. As a consequence, conversational agents are likely to be deployed in the near future that are designed to pursue *targeted influence objectives*. Sometimes referred to as the "AI Manipulation Problem," the emerging risk is that consumers will unwittingly engage in real-time dialog with predatory AI agents that can skillfully persuade them to buy particular products, believe particular pieces of misinformation, or fool them into revealing sensitive personal data. For many users, current systems like ChatGPT and LaMDA feel safe because they are primarily text-based, but the industry is already shifting towards real-time voice and photorealistic digital personas that look, move, and express like real people. This will enable the deployment of agenda-driven Virtual Spokespeople (VSPs) that will be highly persuasive through real-time adaptive influence. This paper explores the manipulative tactics that are likely to be deployed through conversational AI agents, the unique threats such agents pose to the epistemic agency of human users, and the emerging need for policymakers to protect against the most likely predatory practices.

## 1 INTRODUCTION

The phrase "epistemic agency" refers to an individual's control over his or her own personal beliefs [1]. When citizens lose their epistemic agency, democracy is threatened because the political establishment can easily deploy propaganda, misinformation, and disinformation that supports authoritarian objectives, interests, or policies [2-5]. Mass media techniques have been used for generations to weaken the epistemic agency of democratic populations, but over the last decade this problem has been amplified by social media technologies that can direct targeted influence campaigns at specific sub-groups [6]. Segmentation and targeting on social media have been shown to drive polarization, promote radicalization, and foster discontent [7,8]. In this context, all new forms of media should be evaluated in their capacity for abuse and misuse, especially when it threatens epistemic agency [9].

It is therefore important to consider the dangers of Conversational AI and its potential use as a highly personalized and interactive form of targeted influence. As used herein, the phrase "Conversational AI" refers to the deployment of automated AI-driven agents that engage individual human users in interactive dialog [10, 11]. When text-based, these systems are generally referred to as chatbots. When combined with natural voice generation and recognition, they are often referred to as virtual agents and can be used in call centers, as voice-based virtual assistants, and other spoken use-cases [12]. When combined with simulated human faces that have an authentic appearance and can express interactive facial sentiments in authentic ways, they are referred to as virtual humans or virtual spokespeople (VSPs), especially when used to represent the specific interests of third parties through natural conversational interactions [13].

Until recently, the prospect that Conversational AI systems could interact with human users through automated real-time dialog that is perceived as coherent, naturally flowing, and context-aware was still a theoretical prospect. But with the deployment of systems like LaMDA from Google and ChatGPT from OpenAI in 2022, it has become apparent that human-level conversations with machines are now within reach. What makes systems like ChatGPT most unique from prior technologies is their ability to interact continuously with a human user, keeping track of evolving conversational context and even asking probing questions to the user to acquire needed clarifications or explanations [14]. Furthermore these technologies will be deployed on a global scale, as Microsoft recently announced that ChatGPT has been integrated into the Bing search engine and Google announced that LaMDA is being deployed in a new tool called Bard [15].

Considering the facts above, it is clear that Conversational AI tools and technologies have made significant advances over the last 12 months and is likely to be deployed widely by major technology companies in the coming years. For these reasons, we must now consider natural conversations between humans and machines as a viable and potentially dangerous deployment vector for targeted influence campaigns. The following sections explore this danger in hope of informing policymakers that Conversational AI, unlike prior forms of mass media, is an *interactive real-time medium* and therefore is susceptible to new abuses that have not been confronted on traditional or social media platforms.

## 2   CONVERSATIONAL AI AND THE LOOMING "MANIPULATION PROBLEM"

The *AI Manipulation Problem* is used to describe human-computer interaction (HCI) scenarios in which an AI-powered system manipulates a human user in real-time in order to achieve *targeted influence objectives* [18, 19], often by performing the following sequence of steps:

- **(i)** Impart real-time targeted influence on an individual user;
- **(ii)** Sense the user's real-time reaction to the imparted influence;
- **(iii)** Adjust influence tactics to increase persuasive impact on user.
- **(iv)** Repeat steps i, ii, iii to gradually maximize influence in real-time.

This may sound like an abstract series of computational steps, but we humans usually just call it — a conversation. After all, when a salesperson wants to influence a customer, an effective approach is often to speak directly with the target and adjust the conversational arguments in real-time when confronted with resistance or hesitation, gradually maximizing the persuasive impact. The danger is that Large Language Models (LLMs) have advanced so rapidly in recent months, conversational AI systems can now engage in flowing dialog that perform the same manipulative steps. Of course, there are many positive applications that justify the development and deployment of conversational AI systems, but we must also consider the significant danger of misuse for targeted coercion, persuasion, and influence.

Whether we like it or not, interactive Conversational AI systems can now be designed and deployed that draw users into seemingly casual dialog while pursuing *targeted influence objectives* through real-time feedback control [13, 18]. To help policymakers appreciate how different this is from traditional forms of media-based influence, potentially having a much larger impact on epistemic agency, we can describe the risks in the context of feedback control (see Figure 1).

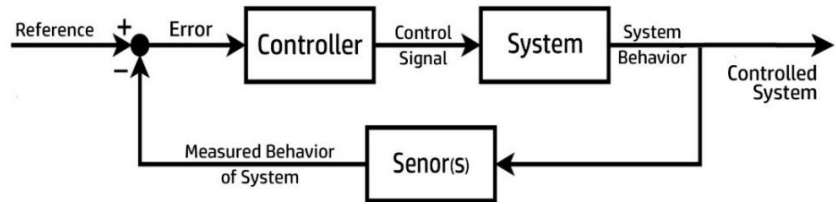

Figure 1: Standard Diagram of Feedback Control System

Control Theory (CT) is the engineering discipline that formalizes how real-time feedback loops can be used to guide the behavior of any system towards a specified goal. The classic example is a thermostat in a house. A homeowner sets a temperature goal and if the house falls below the goal, a controller turns up the heat. If the temperature rises too much, the controller reduces the heat. When working properly, the thermostat keeps the house close to the specified objective.

Referring to Figure 1, the *System* being controlled in the heating example is a house, the *Sensor* is a thermometer, and the *Controller* is a thermostat that modulates the heat as needed. An input signal called the *Reference* is the desired temperature goal. The goal is compared to the actual temperature in the house (i.e., the *Measured Behavior*) and the difference is fed into the controller which then determines how to adjust the heat. This creates a real-time *feedback loop* that continually detects behavior (i.e.., temperature) and imparts influence (i.e., heat).

While a controller can be as simple as a thermostat, it can also be extremely complex. For example, self-driving cars use AI-based controllers to navigate traffic, achieving goals in rapidly changing environments. When considering the use of Conversational AI to impart influence on a user, similar feedback-control methods could be abused. Instead of a simple thermostat that turns up or down the heat as needed, an AI Agent that engages the user in real-time dialog could modulate its persuasive tactics based on the *Measured Behavior* of the system, which in this case are the real-time reactions of the user during the conversation. The *Reference* signal, instead of being a temperature goal, becomes the *influence objectives* of a third party such as a *corporate sponsor* or *state actor*. This yields the manipulative control system of Figure 2 below.

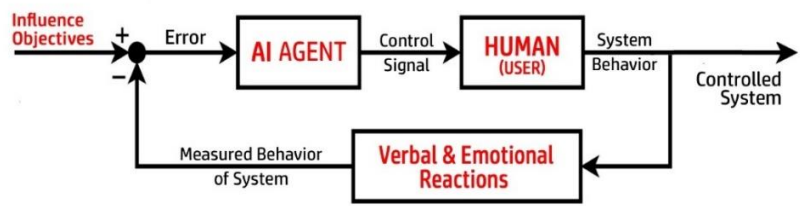

**Figure 2:** Manipulative Control System using Conversational Agents

Such AI-driven control systems are likely to be highly skilled at achieving persuasive goals. After all, the large tech platforms that deploy such conversational agents will likely have extensive personal data about target users including their interests, hobbies, values, and background. This data can be used to craft dialog that is custom designed to draw the target into conversation.  Once engaged, the target will provide verbal responses that are assessed by the AI controller in real-time, enabling the conversational agent to adjust its arguments to counter any resistance or hesitation. In addition, the feedback control system will likely analyze the target's emotional reactions in real-time, processing their vocal inflections for emotive content. And finally, these systems will likely use webcams to process the target's facial expressions, eye motions, and even pupil dilations in real time – all of which infer their emotional reactions at every moment during the interactive dialog, enabling the controller to continuously adjust conversational tactics for maximized impact.

Some argue that current systems like ChatGPT are not dangerous because they're text-based, but the industry is already shifting to real-time voice and photorealistic digital personas that look, move, and express like real people. This will enable the deployment of agenda-driven Virtual Spokespeople (VSPs) that are highly impactful and convincing [24]. Still, some argue that the manipulatory risks are not new threats, as human salespeople already do the same thing, reading emotions and adjusting tactics. Unfortunately, AI systems are likely to be far more perceptive than human representatives. For example, AI systems can detect micro-expressions on human faces that are far too subtle for human observers [20]. Similarly, AI-systems can read faint changes in human complexion known as *facial blood flow patterns* and subtle changes in *pupil dilation* to assess emotions in real-time [21,22].

In addition, these platforms are likely to store data during conversational interactions tracking over time which types of arguments and approaches are most effective on each user personally. For example, the system could learn whether a target user is more easily swayed by factual data, emotional appeals, or by playing on their insecurities. In other words, these systems will not only adapt to your real-time emotions, but they will also get better and better at "playing you" over time, learning how to draw you into conversations, guide you to accept new ideas, and convince you to buy things you don't need, believe things that are untrue, even support extreme policies or politicians that you'd naturally reject.

Some argue that AI agents will not be as skilled at persuasion as human representatives. Unfortunately, without regulatory protections, these AI systems are likely to be trained on sales tactics, human psychology, and other forms of human persuasion. In addition, recent research shows that AI systems can be highly strategic. In 2022, DeepMind used a system called DeepNash to demonstrate for the first time that an AI could learn to bluff human players in games of strategy, sacrificing game pieces for the sake of a long-term win [23]. From that perspective, consumers will be extremely vulnerable to persuasion when facing off with AI-powered conversational agents designed for strategic influence.

## 3   CONCLUSIONS AND RECOMMENDATIONS

For the reasons described above, regulators and policymakers should consider conversational AI a significant threat to epistemic agency [12, 20, 25].  While current "influence campaigns" on social media are analogous to buckshot fired at broad groups, conversational agents could function more like "heat seeking missiles" that adapt their tactics in real time to maximize impact on individual users. We cannot rely on existing regulatory protections aimed at preventing current abuses on traditional media or social media.  This new influence threat will be different – it will be personalized, interactive, and strategic, and therefore requires different forms of regulation. At a minimum, regulators should consider policies that ban platforms from "closing the loop" around human users for persuasive purposes. For example, platforms could be barred from tracking real-time emotional reactions of target users through vocal inflections, facial expressions, or other biometrics.